# Secure, Scalable and Privacy Aware Data Strategy in Cloud

Vijay Kumar Butte[1], Sujata Butte[*2]

[*]*University of Idaho, Idaho Falls, ID, USA*
[2]`suja2591@vandals.uidaho.edu`
[1]`vkbutte@gmail.com`

*Abstract*— The enterprises today are faced with the tough challenge of processing, storing large amounts of data in a secure, scalable manner and enabling decision makers to make quick, informed data driven decisions. This paper addresses this challenge and develops an effective enterprise data strategy in the cloud. Various components of an effective data strategy are discussed and architectures addressing security, scalability and privacy aspects are provided.

## I. Introduction

The world today is seeing exponential growth in the amount of data. This data provides wealth of information that can be used to deliver business and social value. One of the aspects of growing volume, veracity and velocity of data is the challenge of processing and storing the data in a secure, cost effective and efficient manner. An enterprise needs to efficiently process, store large amounts of structured, semi-structured and unstructured data and make data available to data consumers with minimum delay. The businesses need to make quick and informed data driven decisions and adapt to changing market conditions and business dynamics. This is only possible if the data consumers are provided timely, secure and efficient access to data. This gives rise to a critical need to develop a holistic enterprise data strategy and architecture.

Over the last decade, the enterprises have been rapidly adopting cloud and moving the digital assets from on premise to public cloud [1-3]. The enterprises are motivated by the efficient and cost-effective infrastructure provided by cloud service providers. This gives rise to a need for businesses to develop a modern data strategy that is relevant to state of the art cloud technologies.

The data owners, regulators and society in general is more privacy conscious today [4-6]. As the regulators develop new privacy guidelines and regulations, enterprises need to proactively become privacy aware and minimize privacy violation risks through effective data strategy.

The traditional approaches of data strategy and architecture fall short of addressing these challenges and are suboptimal in meeting today's business needs. This paper addresses this gap and develops a holistic enterprise wide data and analytics strategy. The paper focuses on developing a secure, scalable and privacy aware data architecture in the cloud.

The paper is organized as follows, section 2 provides an overview and key component of effective data strategy. Section 3 proposes a holistic data architecture. The architecture considering security, privacy and scalability are provided in this section. Further, the practical aspects of architecture implementation are discussed in this section. This is followed by the conclusion in section 4.

## II. Components of Effective data strategy

An effective data strategy develops an enterprises' vision for collecting, storing, sharing, and usage of its data. The key components of an effective data strategy include people processes and technology. This paper focuses on technology and process aspects. The key aspects include data sources, data transportation and ingestion, data storage and processing, data consumption and analytics, data security and data governance. The figure 1 shows the key components of an effective data strategy.

### A. Data Sources

The data needed for business is obtained from various sources. This includes various databases, enterprise resources, file and object stores, data collected by event collectors, from external applications, log data etc. The source data can be broadly classified into batch data sources and streaming data sources. In batch sources, the data is all processed at regular intervals of time and ingested into the destination for consumption. The streaming data is generated continuously and needs to be processed and consumed as the data becomes available.

*B. Data Transportation and Ingestion*

The data from various sources is securely transported and ingested into cloud data storage for further processing. The data transportation can be achieved in various ways, these include data replication, workflow management and event streaming. In the data replication, the data from source systems is replicated into the cloud data store at regular frequency. The workflow management facilitates streamlining and standardizing the data workflow management. Using the event streaming component, the data from the streaming source is streamed to cloud storage. The event streaming will have both hot and cold modules. The data streaming services will efficiently capture, stream and store data streams. It performs needed transformations and delivers to downstream services. An enterprise needs to approach data transportation and ingestion strategy with due diligence. This includes a well thought out plan for data security and compliance, cost management and speed of delivery. The data migration is carried out with a well planned migration strategy. This includes the discovery phase for identifying the sources that need to be moved to the cloud. This is followed by an assessment phase. In this phase, the enterprise selects optimal destination storage options, develops a secure and efficient migration method and migration tools. This is followed by the migration phase, where the data is moved progressively to the cloud. The migration starts with an initial pilot migration and is continuously improved and fine tuned with new learnings in the process.

*C. Data Storage and Processing*

The data from various sources is received in the cloud. The objective in this layer is to store and process the data such that the end user or data consumer can access the data securely and efficiently and deliver business value quickly. The data is received and processed in various zones. The data quality, data privacy and data security aspects need to be addressed efficiently. Eventually high-quality data is delivered to end users in the form of a purpose-built store. This forms the heart of an effective data strategy. A detailed discussion is carried out in the next section.

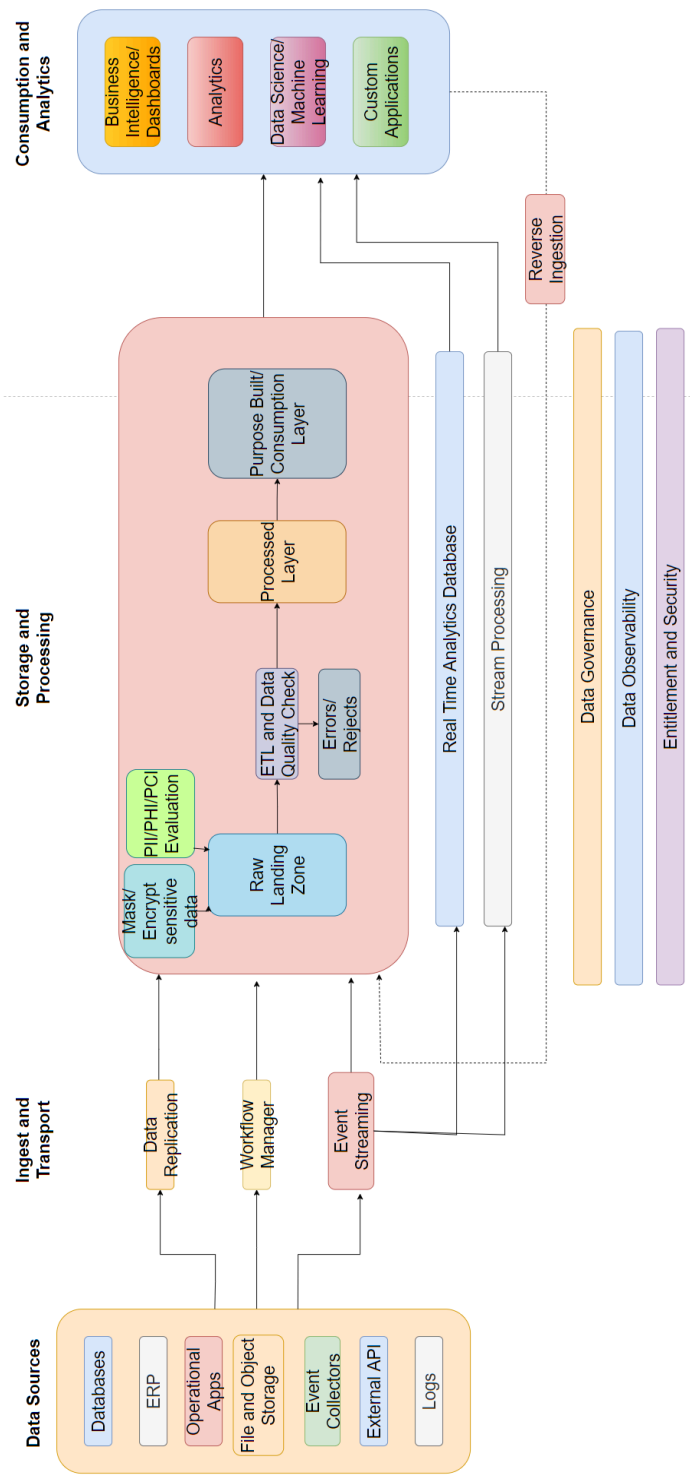

Fig.1 Unified Modern Data Strategy

*D. Data Consumption and Analytics*

The data consumers access the authorized data in a secure manner and use it to generate business value. The data consumers can be business intelligence developers, machine learning engineers, data scientists, application developers etc. It may be noted that these teams may consume the data and generate additional data from their analysis. This data needs to be ingested back to the data lake to make the secondary data generated from data consumers available to the rest of the teams within the organization. This is called reverse ingestion. The reverse ingestion process goes through the same steps as a data source and gets ingested into a data lake.

*E. Data Governance and Cataloguing*

The objective of an effective data architecture is making high quality data available to data consumers in a secure and efficient way. This involves ensuring that the data lake resides in a highly secure and reliable platform. The governing mechanisms ensure that the data is cleaned, processed, protected and classified. This also involves making reliable information about the data assets and its meta data available through effective cataloguing. It may be noted that the governance, risks and compliance aspects are defined and implemented methodologically enterprise wide.

*F. Data Security*

The data security aspect focuses on providing a defined, concrete set of policies and practices that protect the data in the data lake and prevent unauthorized access to the data. This also considers that security policies and patterns are efficient so that they don't not create unneeded burden on data consumers in accessing the authorized data [7-8]. The data security aspects include authorization, encryption, authentication. Authorization component focuses on providing the right level of access to authorized team members based on their roles and personas. The encryption component ensures that the data is scrambled so that in the event of breach the data remains un-inferable. The authentication aspect provides a process to verify the identity of an entity attempting to access the data and determine if this entity has the right level of access permissions.

*G. People*

Learning need analysis

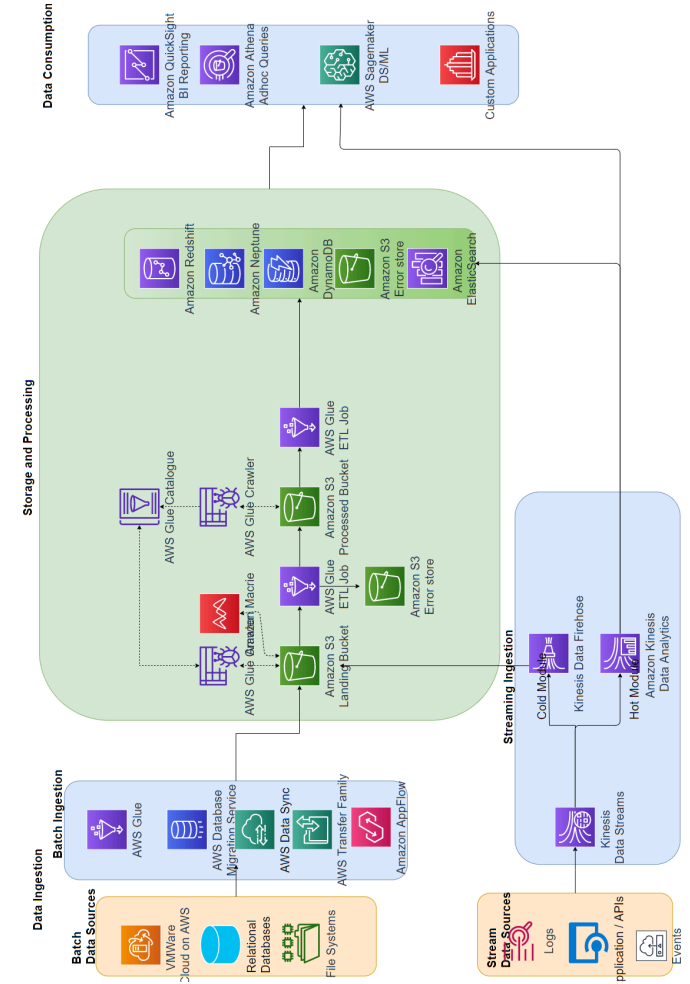

Fig.2 Unified Effective Data Strategy in Cloud

The figure 2 shows unified data strategy in Amazon Web Service (AWS) cloud. The authors are not representing AWS. The batch data from various sources is transported and ingested through various services such as AWS Glue, AWS Database Migration Service, AWS Data Sync, AWS transfer family, Amazon AppFlow etc. The streaming data is handled by AWS Kinesis services. The streaming data is processed using Kinesis Data Stream. Further, the data is sent to Kinesis data Firehose for batch ingestion into the data lake. While the data that need live inference or processing is sent to Kinesis data Analytics for analytics processing. Additionally, the hot module feeds data to other machine learning and live inference needs downstream.

The data from various sources lands in the landing zone Amazon Simple Storage Service (Amazon S3) bucket. AWS Glue is used to process and curate the data as needed by the business. The data quality checks on the raw data is completed using AWS Glue DataBrew. The data is checked for any sensitive information that needs to be encrypted and masked. The masking and encryption events are performed using AWS Glue DataBrew. To further improve the privacy and security aspects Amazon Macie is used to identify and remove any unexpected PII data. Once the data is processed, it is catalogued and registered into the Lake Formation Data catalogue. AWS Glue is used to register and create the catalogues.

Amazon Athena is used to query the data from data lake. The AWS Lake formation verifies the access permission and provides credentials to run the query. The data is received by data consumers and used for further processing. This may involve developing dashboards using Amazon Quicksight, building machine learning models using Amazon Sagemaker or consumption by enterprise business applications etc.

III. DATA LAKE ARCHITECTURE

Data lakes enable efficient storage of large amounts of structured, semi-structured and unstructured data in their raw format [9-14]. The data received in the cloud is systematically processed and stored in the data lake. Taking a zonal approach to data lake architecture provides an effective way to manage and scale data lakes.

*A. Raw landing zone*

The data from various sources reaches the landing zone. In the landing zone, the data is available in its raw format. This forms an optimal node to enforce the data security and governance requirements. The organization can identify the sensitive data fields that need to be accessed only by authorized individuals and applications and encrypt or mask it. Once the data reaches the landing zone, the events to encrypt or mask the specified data items are triggered. The encryptions and masking are performed as needed by regulatory and business requirements.

To further improve the privacy of data, a customized Personally Identifiable Information (PII) detection and removal algorithm can be implemented. This ensures that PII data is identified early in the pipeline and appropriate actions be taken. This may trigger events that inform of any PII violations and remove the sensitive data from the landing zone and move it to the PII error store. The PII error store is then checked to dive deep to understand the sensitivity level and take actions to mask, encrypt or remove it.

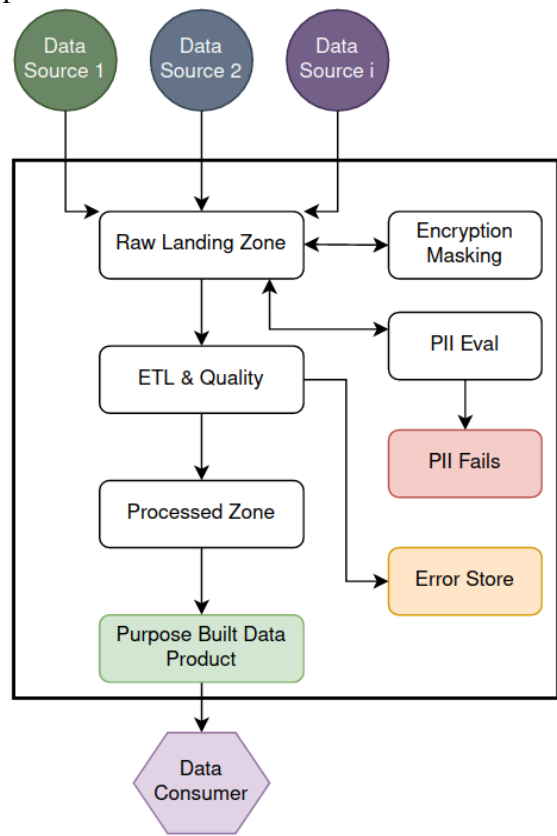

Fig.3 Data lake architecture

*B. ETL and Data Quality*

The data from the raw landing zone is processed for extract transform and loading (ETL) to the processed zone. In this stage the data is checked for generic business quality checks. This involves validating the data for expected quality requirements, removing incorrect or low-quality data and eliminating any unneeded duplications. Any errors are sent to the error store for further analysis. The data quality requirements are defined and tracked. The initial data quality metrics of the data products are documented and updated at regular frequency.

This stage also performs the generic aggregations and produces metadata to describe the dataset.

*C. Data encryption and masking*

Enterprises need to mitigate the risk of exposing sensitive data to unauthorized personnel. This may involve limiting access to personally identifiable information (PII), personal health information (PHI), business sensitive data etc. The enterprises need to identify and categorize and label the data based on its level of sensitivity. Additionally, a logic should be in place to proactively identify the sensitive data that may have not been labelled already.

The businesses come across three types of data sources. They include the sources that do not have any sensitive information, sources that include medium sensitive and non-sensitive information and sources that are highly sensitive. To address this aspect a layered approach to data lake architecture can be adopted. In this, the sensitive data is ingested into a separate highly secure zone within the data lake. Often the data is client side encrypted. The consumption layer for this highly sensitive data also stays within a highly secure zone and accessed only to specific entities for a defined period of time. The data that is classified as not sensitive is directly landed into the raw landing bucket and follows the regular path of data quality checks and subsequent processing.

The data that has partially sensitive data is first landed into an isolated landing zone bucket. The data masking logic is run on this landing bucket. Once new data is received into the isolated area, an event is triggered where the data is matched with the sensitive profile. If any sensitive data is found, the recipe jobs are created to mask the sensitive data items. If no sensitive data is found the process is marked as complete.

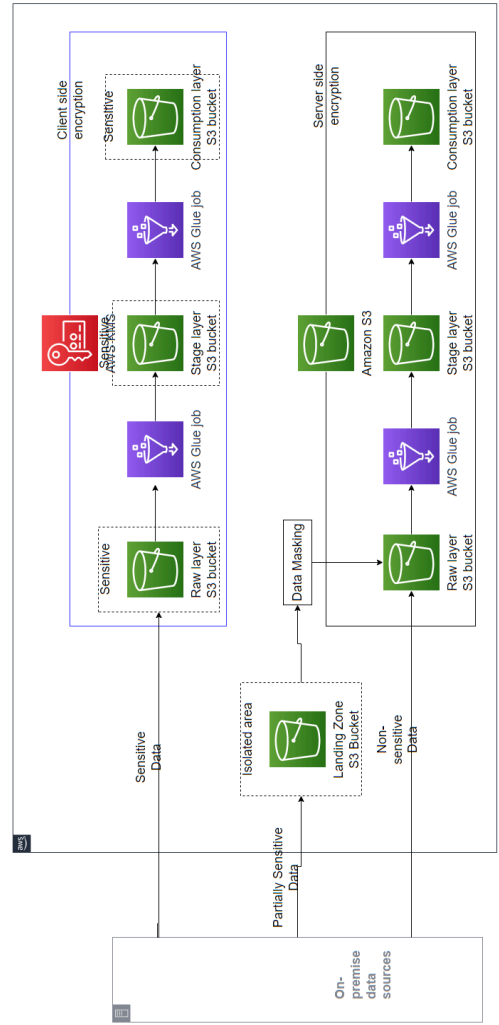

Fig.4 Secure layered data lake architecture

The figure 5 shows the masking architecture in AWS cloud using Amazon Simple Storage Solution (S3), Amazon Glue DataBrew, Amazon EventBridge, AWS lambda and AWS Secrets Manager. Amazon Glue DataBrew is a data preparation tool that is used for data preparation and matching the profile jobs. The AWS lambda is a serverless, event driven compute service used to run custom logic. Amazon S3 allows simple storage of any type of data at scale. AWS step function is a workflow service that allows users to build distributed applications, automate IT and business processes. AWS secrets manager allows users to manage, retrieve and rotate credentials, API keys and other secrets.

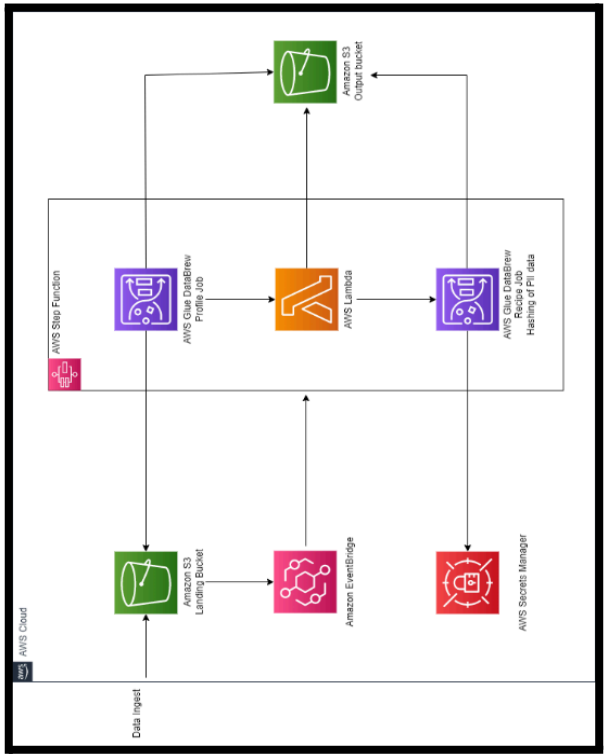

Fig.5 Masking architecture for data lake

*D. PII evaluation*

The data lakes receive a large amount of data both from internal and external sources. The businesses do due diligence to classify and label the data and its level of sensitivity [15]. However, there is a risk of sensitive data passing the classification effort. This may result in highly sensitive data being made available to a large number of users. This may lead to regulatory compliance violations, loss of trust from customers and bad actors acting to create negative consequences for the business. Hence, it is critical to have an automated check to detect sensitive information. Once the data is landed into the landing zone after the masking is complete, a blind check is done to detect any potential sensitive data.

The figure below shows the PII data inspection architecture in AWS cloud using Amazon S3, Amazon Macie, Amazon EventBridge, AWS lambda and Amazon Simple Notification Service (SNS). Amazon Macie is an AWS managed data security and privacy service that uses machine learning and pattern matching to discover and protect any sensitive data. Amazon SNS is a managed messaging service for application to application and application to person communication and enables users to send notifications. Once the data is landed into the landing zone, a data discovery algorithm is triggered. This finds sensitive data for PII, PHI and PCI. If medium sensitive data is found, it will trigger the logic to notify the concerned team. A manual evaluation is done to take remedial actions. This may include updating the enterprise data sensitivity classification document and the masking logic. If highly sensitive data is found, the lambda function is triggered to notify the concerned team and delete or move the sensitive data from the landing bucket into the PII error store. This minimizes the probability of PII information leaking to downstream processes.

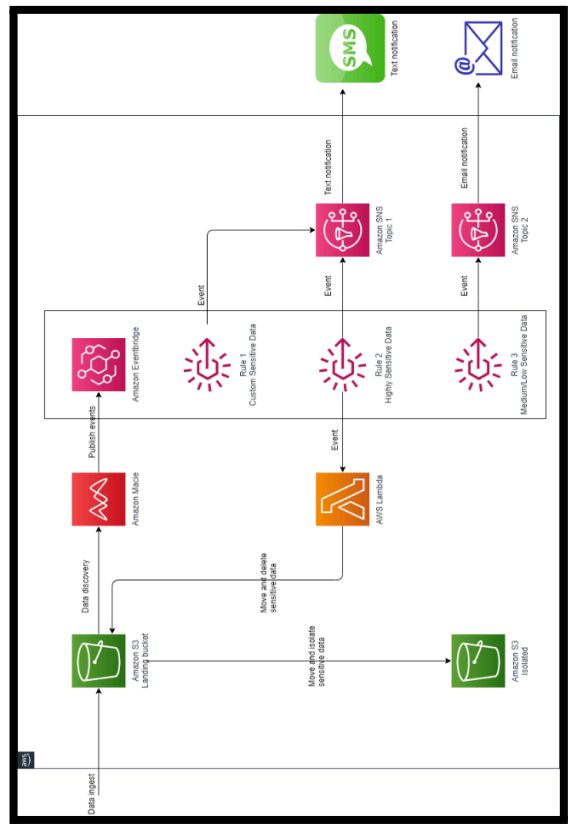

Fig.6 PII detection and removal architecture

*E. Processed Zone*

The processed layer containing the data is sourced from the landing zone. In this zone the data is stored for long term usage. This serves as a single source of trusted data for downstream processes and applications. The data in this zone is processed to meet the overall business needs. This results in the data that is enriched, indexed data with relevant metadata.

*F. Data Product Layer*

The data product layer sources data from the processed layer and builds data products as needed by the data consumer. The second phase of compliance and quality checks can be enforced to ensure that the high-quality data is made available to business applications and teams. These data products enable a specific set of business applications and advanced analytics. This enables reusable data products that can be sourced by various applications and teams within and outside the domain. The data product should be prepared for secure and efficient sharing of data across enterprise. This includes providing various details such as, data product name, meta data (ownership, schema, semantics and security), data quality, details of the use cases for which the data product is suitable, way to access the data product etc.

*G. Data Consumption Layer*

The consumption layer has the tools and services needed by data consumer groups. These may include Business Intelligence (BI) developers who use the BI dashboards that are consumed by business end users. These also include machine learning platforms where data scientists use the data from a purpose-built layer or the processed layer to build machine learning models. Further the data in the purpose-built layers can also be made available to any internal business or appropriate third-party application.

*H. People*

People form a key component of an effective data strategy. The commitment from senior leadership is essential to procure needed headcount and financial resources. The data strategy team should have diverse representation from various stakeholder teams. It is critical to clarify the roles and responsibilities and set clear expectations from various stakeholders. The teams should be provided with needed training and enabled with learning resources. Additionally, it is critical to establish an effective communication plan between various stakeholders.

IV. CONCLUSIONS

The paper discussed key challenges businesses face in the growing data domain. To address these challenges, it provided an effective data strategy in the cloud. The key aspects of effective data strategy such as security, privacy and scalability were addressed by providing architectures and patterns for practical implementation.

People form a key component of an effective data strategy. The commitment from senior leadership is essential to procure needed headcount and financial resources. The data strategy team should have diverse representation from various stakeholder teams. It is critical to clarify the roles and responsibilities and set clear expectations from various stakeholders. The teams should be provided with needed training and enabled with learning resources. Additionally, it is critical to establish an effective communication plan between various stakeholders.